\documentclass[10pt,twocolumn,letterpaper]{article}

\usepackage{cvpr}
\usepackage{times}
\usepackage{epsfig}
\usepackage{graphicx}
\usepackage{amsmath}
\usepackage{amssymb}
\usepackage[normalem]{ulem}
\useunder{\uline}{\ul}{}
\makeatletter
\newcommand\footnoteref[1]{\protected@xdef\@thefnmark{\ref{#1}}\@footnotemark}
\makeatother

\usepackage[breaklinks=true,bookmarks=false]{hyperref}

\cvprfinalcopy 

\def\cvprPaperID{18} 

\ifcvprfinal\fi
\begin{document}

\title{Natural Image Noise Dataset}

\author{Benoit Brummer\\
Université Catholique de Louvain\\
Louvain-la-Neuve, Belgium
{\tt\small benoit.brummer@student.uclouvain.be}
\and
Christophe De Vleeschouwer\\
{\tt\small christophe.devleeschouwer@uclouvain.be}
}

\maketitle

\begin{abstract}

Convolutional neural networks have been the focus of research aiming to solve image denoising problems, but their performance remains unsatisfactory for most applications. These networks are trained with synthetic noise distributions that do not accurately reflect the noise captured by image sensors. Some datasets of clean-noisy image pairs have been introduced but they are usually meant for benchmarking or specific applications. We introduce the Natural Image Noise Dataset (NIND), a dataset of DSLR-like images with varying levels of ISO noise which is large enough to train models for blind denoising over a wide range of noise. We demonstrate a denoising model trained with the NIND and show that it significantly outperforms BM3D on ISO noise from unseen images, even when generalizing to images from a different type of camera. The Natural Image Noise Dataset is published on Wikimedia Commons such that it remains open for curation and contributions. We expect that this dataset will prove useful for future image denoising applications.

\end{abstract}

\section{Introduction}

Photographic image noise occurs as a camera sensor's ISO sensitivity increases to capture an image faster than it would in ideal conditions (``base ISO" sensitivity).\footnote{We often make references to ISO noise because increased ISO sensitivity is the main cause of noise, but it should be noted that there are other factors affecting the magnitude of noise acquired by the image sensor.} A fast shutter speed is often necessary even though there is insufficient light, for instance with handheld photography where a slow shutter speed results in blur caused by the camera shake, or when a dynamic subject results in motion blur. Increasing the ISO setting is akin to linearly amplifying the value measured on each sensor cell. A small initial value that is amplified is less accurate and more prone to errors; this amplified value in turn makes up photographic noise. 

Denoising is typically seen as the inverse problem of recovering the latent clean image from its noisy observation \cite{rednet}. The use of deep learning to solve the denoising problem by directly generating the latent clean image, or in some cases recreating the noise and subtracting it from the observed image \cite{dncnn}, has been investigated. However, while the synthetic results show state-of-the-art performance, testing on real data indicates that neural network-based solutions do not exceed the performance offered by BM3D \cite{darmstadt}. It appears that neural networks simply learn the applied noise distribution and that ISO noise may involve additional transformations such as color distortions and loss of detail.

Some specialized work has shown that neural networks obtain state-of-the-art performance when trained with real data \cite{learningtoseeinthedark}\cite{microscopynoise}. We sought to assess the potential of deep learning applied to the denoising problem by expanding on this previous work through a dataset of images produced with various levels of ISO noise. This dataset can be used to train neural network models for general purpose denoising of high quality images.

Our work introduces an open dataset of DSLR-like\footnote{We define a DSLR-like camera as one produced with an APS-C (25.1x16.7 mm) or larger sensor such as those present in most DSLR and mirrorless cameras} image sets with various levels of real noise caused by the digital sensor's increased ISO sensitivity. The dataset is large enough to be used for training and varies in content in order to model a great variety of scenes. Each scene was captured at multiple noise levels, with an average of 6 images per set, such that a model may be trained for blind denoising on the base ISO as well as beyond the highest ISO value of the camera by feeding it crops that have a random noise value.

The images in a set are all pixel-aligned. Some of the scenes include multiple ground-truth images which may be sampled at random during training; these would prevent the model from learning to reconstruct the random noise it has seen on one ground-truth, thus making it more difficult to overfit the noise. Overexposed areas are avoided in the ground-truth images because details are lost when the sensor is saturated and this could potentially give an advantage to higher ISO images in which the sensor is not necessarily saturated (notably in ISO-invariant cameras and high ISO pictures that are brightened using software).

We trained a U-Net model with the dataset and validated its denoising performance over increasing ISO on a test set taken with the same Fujifilm X-T1, as well as the generalization on a separate subset with scenes captured using a Canon EOS 500D. The dataset is published in sRGB format on Wikimedia Commons, which is an open-platform that promotes continuous discussion and contribution.
\section{Related work}
The following works feature datasets made of ground-truth / noisy image sets. The static scenes approach is necessary to directly compare the level of degradation using a loss function such as the structural similarity index (SSIM) \cite{ssim} or the mean square error (MSE). 
\subsection{Darmstadt Noise Dataset}
The Darmstadt Noise Dataset (DND) \cite{darmstadt}, containing 50 pairs of noisy-clean images from four cameras, was developed for the purpose of validating denoising algorithms using real data. Synthetic noise is typically used to train and test models, but it had been unclear whether the reported synthetic results translated to real improvements. Plötz and Roth showed that many modern denoising methods do not perform well on real data and that BM3D \cite{bm3d}, which was published in 2006, remains one of the best performing methods \cite{darmstadt}. RENOIR \cite{renoir} is a similar dataset that was published prior to the DND; however, Plötz and Roth noted spatial misalignments that reduced its effectivity. We have additionally found that the light sometimes differs between images in the same scene and that some photographs exhibit significant raw overexposure.
\subsection{Learning to See in the Dark}
See-in-the-Dark (SID) \cite{learningtoseeinthedark} is an image noise dataset that is large enough for training and, to our knowledge, was used in the first successful attempt at denoising images using real image noise. This dataset focused on very low-light photography where the camera-generated JPEG appears black. The authors used a U-Net network architecture to create an end-to-end RAW-to-JPEG pipeline that produces realistic colors, improving on standard processing and BM3D denoised images which still suffer from color bias at high ISO. Our work differs from SID in that we aimed to train a general purpose (``blind") denoiser rather than one that handles a specific condition, such as extremely low light images. We chose to work in sRGB space because handling the whole RAW-to-sRGB pipeline removes some information which may otherwise be useful to the author during development. Moreover, one dataset can then be used with different types of color filter arrays.
\subsection{Smartphone Image Denoising Dataset}
The Smartphone Image Denoising Dataset (SIDD) \cite{sidd} is comprised of 10 scenes * 5 cameras * 4 conditions * 150 images, totalling 30000 images. This dataset aims to address the problem of smartphone image denoising, where the small sensor and aperture size causes noticeable noise even in pictures taken at base ISO. Further processing is thus applied to create ground-truth images out of many images. This method of creating ground-truth images is not entirely relevant for denoising images captured with larger sensors because a single image taken at base ISO on a DSLR-like camera is clean enough to work as ground-truth for training purposes.
%

\section{Dataset}
Here we outline the physical setup required to capture image sets for the NIND, summarize its content, explain the software processing and validation requirements, and describe its publication aspects such that others that wish to do so may also contribute.
\subsection{Capture}
We captured several images per static scene; at least one ground-truth taken with the camera's lowest ISO setting and several images taken with increasing ISO settings and consequent decreasing shutter speed in order to match the original exposure value. Scenes were captured using a camera affixed to a tripod and controlled with a wireless remote control to avoid shifting the setup position. We ensured that the ground was stable, wind would not cause any change in the scene, lighting did not vary between shots, and no area was overexposed in the ground-truth images. Overexposure occurs when the sensor is saturated; on a ground-truth image this would potentially benefit the high ISO images because less light is captured with a faster shutter speed, therefore, the higher ISO images may not be overexposed and thus may contain detail that is not present in the overexposed ground-truth. The aperture remained the same on all shots and the focus was set manually so that it would not automatically adjusted for each frame.

A base ISO image (ISO200 on the Fujifilm X-T1) was always taken at least once, along with the camera's highest ISO setting (ISO6400 on the X-T1). Several images were taken with different intermediate ISO values such that the ISO settings varied across each scene. We often also took images that we categorized as ``High ISO," which consisted of the highest ISO value and increased shutter speed. ``High ISO" images result in dark frames which are then correctly exposed using software. We often tried to match shutter speeds that would be useful to denoise, such as 1/60s for handheld photography, 1/15s for devices equipped with optical image stabilization, and 1/1000s or faster for high-speed photography. We ensured that every ISO value was well represented in order to train models effective at blind denoising. On average, six images were produced per scene and some scenes featured multiple ground truth images which could be used in training to help prevent over-fitting.
\subsection{Content}
\begin{figure*}
\centering
  \includegraphics[width=1\linewidth]{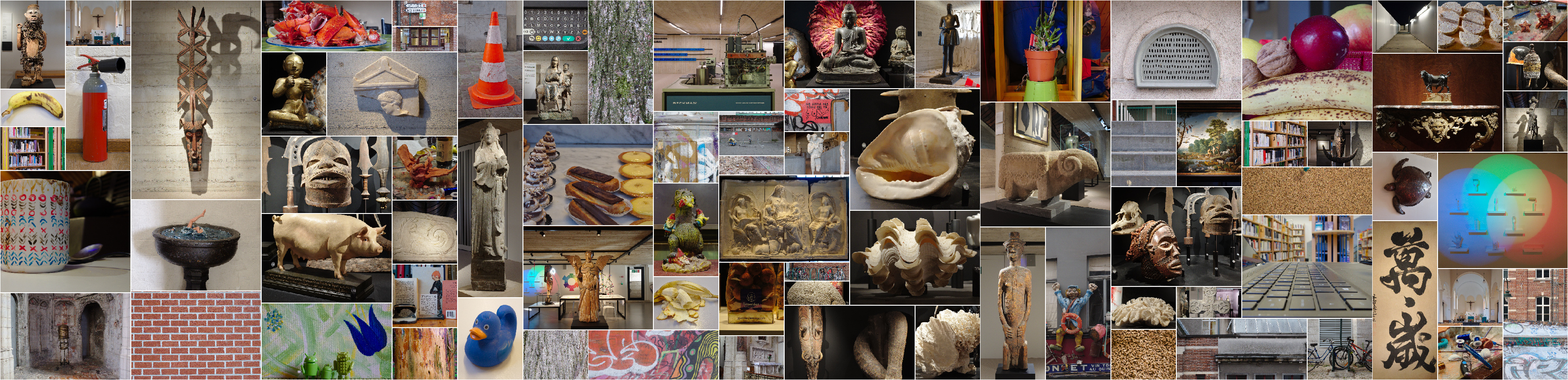}
   \caption{Sample ground-truth images from the Natural Image Noise Dataset.}
\vspace{-0.2cm} 
\label{fig:sampleimg}
\end{figure*}

\begin{table*}[]
\centering
\resizebox{1\linewidth}{!}{
\begin{tabular}{lllllllllllllllllll|ll}
{\ul ISO value} & {\ul 100} & {\ul 200} & {\ul 250} & {\ul 320} & {\ul 400} & {\ul 500} & {\ul 640} & {\ul 800} & {\ul 1000} & {\ul 1250} & {\ul 1600} & {\ul 2000} & {\ul 2500} & {\ul 3200} & {\ul 4000} & {\ul 5000} & {\ul 6400} & {\ul High} & {\ul Scenes} & {\ul Images} \\
{\ul Fujifilm X-T1}      &           & 105       & 11        & 8         & 18        & 13        & 16        & 25        & 16         & 9          & 12         & 9          & 19         & 25         & 17         & 19         & 91         & 123        & 90         & 536          \\
{\ul Canon C500D}     & 14        & 10        &           &           & 10        &           &           & 10        &            &            & 9          &            &            & 11         &            &            &            & 10         & 11         & 74           \\ \hline
{\ul Total}     & 15        & 116       & 11        & 8         & 29        & 13        & 16        & 36        & 16         & 9          & 22         & 9          & 19         & 37         & 17         & 19         & 91         & 133        & 101        & 616         
\end{tabular}%
}\caption{Dataset content}
\vspace{-0.2cm} 
\label{tableDSContent}
\end{table*}

Many objects were captured in museums where subjects are plentiful (albeit we had to be mindful of copyright restricted material) and for which a denoising application would be highly relevant because indoor handheld pictures require a high ISO sensitivity. However, initial tests found that using images taken only indoors did not provide the variety needed to create a model that generalized well across all conditions, as natural colors were sometimes off in outdoor and brightly lit applications. We thus captured natural objects with vibrant colors (such as food items and plant-life) as well as outdoor scenes where the shutter speed could be taken as fast as 1/13000s using a digital shutter. We made an effort to include some text because it is prevalent, yet we expect a model would not be able to guess how to reconstruct it (Figure \ref{fig:text} shows the resulting denoised text), and we tried to make the images pleasant to look at in order to enhance the time users would spend looking at them. Most of the NIND images were captured on a Fujifilm X-T1 mirrorless camera, which uses a 23.6 x 15.6 mm X-Trans sensor. A Canon EOS 500D DSLR camera, featuring a 22.3 x 14.9 mm standard Bayer sensor, was used to capture images that could be used to validate the generalization. The content of the dataset is summarized in Table \ref{tableDSContent} and a subset of the X-T1 pictures are shown in Figure \ref{fig:sampleimg}.
\subsection{Software processing}
We processed the dataset images using darktable \cite{darktable} (an open source development software) for raw-to-sRGB development. Our development steps are similar to those we would apply to a standard picture but no sharpening was applied, as this greatly amplifies noise and is typically applied last in the pixel pipeline (we can expect users to apply sharpening to the generated clean image without any perceptible loss). We used darktable's automatic exposure mode to match a fixed percentile target on the histogram and calculate the required exposure compensation for all images in a set. Likewise, we ensured that the white balance was identical and all development steps were copied over to the entire scene. (The exposure percentile and white balance settings are fixed within a scene but vary across different scenes.) The raw overexposed indicator was used to verify that no overexposed areas were present in the base ISO image and if any were detected then it was cropped out or the scene was discarded. The images were visually inspected to detect slight variations in light, the introduction of foreign objects such as insects, or any movement, which also resulted in cropping or discarding images. The ground-truth images must be at least as sharp as their noisy counterparts; this is sometimes not the case due to slight movements on longer exposures. The remaining images were saved in either high-quality (98 to 100) 8-bit JPEG or lossless 16-bit PNG.

The last step of development is to use Hugin's align\_image\_stack tool \cite{hugin} to ensure that all images in a set are perfectly pixel-aligned. The tool will usually return the same image size as the input, in which case the whole image set can safely be used. When a difference is detected then the tool will automatically align the set and we visually analyzed whether the result was acceptable or the movement caused a change in perspective, in which case the outlier images were discarded. Some noisy images cannot easily be matched to the scene; possible solutions are to denoise these images in order to check the alignment or to take a cleaner image afterward and assume that the middle images are consistent with the previous and next ones.
\subsection{Publishing}
The Natural Image Noise Dataset is published on Wikimedia Commons ( \url{https://commons.wikimedia.org/wiki/Natural_Image_Noise_Dataset} ), an online repository of free-use images and other digital media. Wikimedia Commons hosts media content for all Wikimedia projects and its scope is limited only by the content having some educational value (broadly meaning ``providing knowledge; instructional or informative"). As such, we believe it is a fitting platform for the publication of a research dataset.

One key advantage of using Wikimedia Commons is its collaborative aspect. Anyone is allowed to add images to the dataset, modify existing images (for example to fix a spatial misalignment), and discuss the content (through the discussion page provided for each file, category, and the dataset itself). The collaborative aspect also includes a ``Quality images candidates" page \cite{qic} where users assess the technical quality of a submitted image and may promote it to a ``Quality image" standing. Many of the ground-truth images have gone through this process and were promoted through human assessment. The same process was also used to validate the trained model, which ended up in a positive assessment since even the denoised dynamic ISO6400 picture presented in Figure \ref{fig:visualpigeons} was among the promoted images.

On the technical side, Wikimedia Commons preserves images as they are uploaded; JPEG images are not recompressed, 16-bit lossless TIFF and PNG images are allowed, and the metadata is kept. Thumbnails are generated and images may be visualized before being downloaded in full resolution, and the download may include a select subset instead of the whole dataset. A customizable download script is provided on the dataset's page for convenient retrieval. Even though files can be overwritten, every file uploaded on Wikimedia Commons is kept forever therefore specific snapshots of the dataset can be made by including the files' revision in the download script and getting a specific version ID (or commit hash) of the download script.

It is also important to note the usefulness of the data outside of a denoising dataset context. Many images, such as those depicting artifacts displayed in churches and museums, have encyclopedic value and the ground-truth images present in our dataset are of higher quality than most previously available images depicting such artifacts. By publishing the dataset on Wikimedia Commons, these ground-truth images may be used in Wikipedia articles directly\footnote{E.g. \href{https://en.wikipedia.org/wiki/Bobo_people}{Bobo people}, \href{https://en.wikipedia.org/wiki/Bombardment_of_Brussels}{Bombardment of Brussels}, \href{https://en.wikipedia.org/wiki/Dengese_people}{Dengese people}}. This may be a motivating factor to those wishing to contribute.
\section{Results}
This section describes the suggested use of the dataset. Various network configurations are described in the Model subsection, Usage explains our suggested handling of the dataset, along with some of the tools we provide for this purpose, and we show preliminary results.
\subsection{CNN Denoiser}
\subsubsection{Model}
We initially trained a DnCNN \cite{dncnn} model on our data. This model attained satisfying performance when trained to model the latent clean image instead of modeling the noise. It was further improved by using a convolution filter size of 5x5 instead of 3x3. The second architecture tested was a Red-Net \cite{rednet} with 22-layers and a filter size of 5x5. This model obtained very good performance, albeit with an impractical runtime and memory use. We settled on a U-Net \cite{unet} architecture which provides slightly better performance with significantly lower runtime and memory use.
\subsubsection{Usage}
The dataset is cropped in advance to speed up loading times. A crop size of 128x128 pixels was found to work well for training and larger crops did not significantly affect performance. We found that the border is most often corrupted in a U-Net model. Some network architectures perform better when required to learn to model entire crops down to their border, but it typically takes a lot of resources to get to that point and the result often still shows a grid pattern when the crops are stitched back together. We use a ``useful crop size" that is 0.75 the size of the actual crop size so that only the central part of a crop is used for stitching as well as in scoring. A script that crops the dataset in such overlapping blocks is provided for this purpose.

An epoch consists of training the model on every crop of any ISO value for every scene, that is, a random ISO value is fed every time a crop is loaded, therefore it takes several epochs for the model to train on all of the available data. The ground-truth is also selected randomly when multiple ones are available and basic data augmentation (rotation and/or translation) is performed.
\subsection{Experimental results}

We compared the performance obtained with the following methods:

\begin{enumerate}
\itemsep0em 
  \item\label{exp:nindxt1} U-Net trained on NIND (X-T1 subset):\newline
  This model encompasses the main part of our dataset.
  \item\label{exp:sidd} U-Net trained on SIDD (320 provided image pairs):\newline
  Compare the performance obtained using our dataset with 320 images from the SIDD (Smartphone Image Denoising Dataset) \cite{sidd} which were made available for the 2019 NTIRE denoising challenge.
  \item\label{exp:bm3d} BM3D \cite{bm3d} ( \cite{bm3d-gpu} implementation) with $\sigma$ = \{5, 10, 20, 30, 40, 50, 60, 70, 80, 90, 93, 95, 97, 99\}\footnote{\label{sigmanote}We test every $\sigma$ value mentioned in Methods \ref{exp:bm3d} and \ref{exp:nindart} and report the value which yields the highest SSIM for each test image.}:\newline
  BM3D has been ubiquitously used as a reference in non-learning based image denoising.
\end{enumerate}
In addition to the aforementioned reference methods, we consider the following experiments:
\begin{enumerate}
\itemsep0em 
\setcounter{enumi}{3}
  \item\label{exp:nindall}   U-Net trained on NIND (dataset composed of the union of X-T1 and C500D training scenes; 89.5 \% and 10.5 \%, respectively):\newline
  When tested on a C500D image, this method can be compared with the first reference method to determine whether training on images acquired with the test image sensor helps or not, thereby assessing the generalization capabilities of our reference model to different sensors.  In addition, it shows whether adding data from a different sensor negatively affects performance.
  \item\label{exp:nindsidd} U-Net trained on the union of NIND:X-T1, NIND:C500D scenes) and SIDD (320 pairs); 20.5 \%, 2.4 \%, and 77.1 \% respectively:\newline
  This model shows the performance impact of adding a wildly different type of noise to the training data.
  \item\label{exp:nind6400} U-Net trained on NIND (X-T1 subset, ISO6400 noise only instead of random ISO sample):\newline
  This experiment tests whether a model trained for blind denoising performs significantly worse than one trained for a specific ISO value.
  \item\label{exp:nindart} U-Net trained on NIND (X-T1 subset) with artificial gaussian noise added to the ground-truth. $\sigma$ = \{[1,55], [1,60], [1,80], [1,95]\}\footnoteref{sigmanote}:\newline
  This experiment compares the performance obtained by a model trained on our real data to the widely applied approach of applying synthetic gaussian noise to clean images.
  \item\label{exp:nindmakenoise} U-Net trained to reconstruct the noise on NIND (X-T1 subset):\newline
  We applied the residual learning strategy proposed in \cite{dncnn} by training a model that reconstructs the noise and subtracts it from the image.
  \item\label{exp:nindred} Red-Net \cite{rednet} trained on NIND (X-T1 subset):\newline
  This uses the same data as Method \ref{exp:nindxt1} with a different network architecture

\end{enumerate}

\begin{table*}[!htbp]
\centering
\resizebox{1\linewidth}{!}{
\begin{tabular}{l|llllllllll}
{\ul ISO value}                      & {\ul ISO200}   & {\ul ISO250}   & {\ul ISO500}   & {\ul ISO2500}  & {\ul ISO4000}  & {\ul ISO6400}  & {\ul High ISO} \\ \hline
Number of images                     & 5              & 3              & 2              & 2              & 2              & 5              & 9              \\ \hline
Noisy                                & \textbf{1.000} & 0.907          & 0.853          & 0.784          & 0.687          & 0.578          & 0.311          \\ \hline
NIND:X-T1 (U-Net)                     & 0.949          & \textbf{0.929} & \textbf{0.920} & \textbf{0.912} & \textbf{0.900} & \textbf{0.893} & \textbf{0.851} \\ \hline
SIDD (U-Net)                          & 0.906          & 0.907          & 0.904          & 0.864          & 0.882          & 0.860          & 0.814          \\ \hline
BM3D                                 & 0.941          & \textbf{0.925} & 0.913          & 0.875          & 0.870          & 0.852          & 0.785          \\ \hline
NIND:X-T1+C500D (U-Net)               & 0.949          & \textbf{0.929} & \textbf{0.920} & \textbf{0.912} & \textbf{0.899} & \textbf{0.893} & \textbf{0.851} \\ \hline
NIND:X-T1+C500D + SIDD (U-Net)        & 0.947          & \textbf{0.928} & \textbf{0.919} & \textbf{0.910} & \textbf{0.898} & \textbf{0.892} & \textbf{0.850} \\ \hline
NIND:X-T1 ISO6400 only (U-Net)        & 0.919          & 0.915          & 0.911          & \textbf{0.907} & \textbf{0.901} & \textbf{0.894} & 0.821          \\ \hline
Reconstruct noise on NIND:X-T1 (U-Net)      & 0.950          & \textbf{0.926} & 0.914          & 0.901          & 0.876          & 0.840          & 0.664          \\ \hline
Artificial noise on NIND:X-T1 (U-Net) & 0.963          & 0.920          & 0.899          & 0.880          & 0.810          & 0.769          & 0.531          \\ \hline
NIND:X-T1 (Red-Net)                   & 0.940          & 0.923          & 0.915          & \textbf{0.907} & 0.892          & \textbf{0.886} & 0.842          \\ \hline
\end{tabular}%
}\caption{Average SSIM index on 5 NIND:X-T1 denoised scenes (ursulines-building, stefantiek, CourtineDeVillersDebris, MuseeL-Bobo, ursulines-red). The best performing models (to within two significant digits) are marked in bold.}
\vspace{-0.2cm} 
\label{tableNINDXT1}
\end{table*}

\begin{table*}[!htbp]
\centering
\resizebox{1\linewidth}{!}{%
\begin{tabular}{l|lllllll}
{\ul ISO value}                            & {\ul ISO100}   & {\ul ISO200}   & {\ul ISO400}   & {\ul ISO800}   & {\ul ISO1600}  & {\ul ISO3200}  & {\ul High ISO} \\ \hline
{Noisy}                                & \textbf{1.000} & 0.814          & 0.754          & 0.660          & 0.550          & 0.401          & 0.172          \\ \hline
{NIND:X-T1 (U-Net)}                     & 0.911          & \textbf{0.901} & \textbf{0.898} & \textbf{0.896} & \textbf{0.893} & \textbf{0.887} & \textbf{0.868} \\ \hline
{SIDD (U-Net)}                          & 0.894         & 0.892           & 0.890          & 0.889          & 0.885  & 0.876          & 0.850          \\ \hline
{BM3D}                                 & 0.921          & 0.890          & 0.884          & 0.877          & 0.871          & 0.860          & 0.813          \\ \hline
{NIND:X-T1+C500D (U-Net)}               & 0.911          & \textbf{0.901} & \textbf{0.899} & \textbf{0.896} & \textbf{0.894} & \textbf{0.888} & \textbf{0.872} \\ \hline
{NIND:X-T1+C500D + SIDD (U-Net)}        & 0.912          & \textbf{0.901} & \textbf{0.899} & \textbf{0.896} & \textbf{0.893} & \textbf{0.888} & \textbf{0.871} \\ \hline
{NIND:X-T1 ISO6400 only (U-Net)}        & 0.899          & \textbf{0.897} & \textbf{0.896} & 0.894          & \textbf{0.892} & \textbf{0.886} & \textbf{0.865} \\ \hline
{Reconstruct noise on NIND:X-T1 (U-Net)}      & 0.908          & 0.895          & 0.887          & 0.875          & 0.855          & 0.807          & 0.617          \\ \hline
{Artificial noise on NIND:X-T1 (U-Net)} & 0.946          & 0.879          & 0.864          & 0.836          & 0.802          & 0.716          & 0.430          \\ \hline
\end{tabular}%
}\caption{SSIM index on NIND:C500D denoised set MuseeL-Bobo-C500D}
\vspace{-0.2cm} 
\label{tableNINDXT1onBoboC500D}
\end{table*}

\begin{table*}[!htbp]
\centering
\resizebox{1\linewidth}{!}{%
\begin{tabular}{l|lllllll}
{\ul ISO value}                       & {\ul ISO100}   & {\ul ISO200}   & {\ul ISO400}   & {\ul ISO800}   & {\ul ISO1600}  & {\ul ISO3200}  & {\ul High ISO} \\ \hline
{\ul \# images}                       & 13             & 9              & 9              & 9              & 8              & 10             & 9              \\ \hline
{\ul Noisy}                           & \textbf{0.954} & 0.766          & 0.707          & 0.619          & 0.501          & 0.380          & 0.220          \\ \hline
{\ul NIND:X-T1 (U-Net)}                & 0.878          & \textbf{0.856} & \textbf{0.854} & \textbf{0.848} & \textbf{0.845} & \textbf{0.834} & \textbf{0.805} \\ \hline
{\ul SIDD (U-Net)}                     & 0.851         & 0.838          & 0.837          & 0.834          & 0.830          & 0.813          & 0.774          \\ \hline
{\ul BM3D}                            & 0.899          & 0.836          & 0.825          & 0.816          & 0.811          & 0.789          & 0.749          \\ \hline
{\ul NIND:X-T1 ISO6400 only (U-Net)}   & 0.857          & 0.847          & \textbf{0.847} & 0.845          & \textbf{0.843} & \textbf{0.834} & 0.802          \\ \hline
{\ul Reconstruct noise on NIND:X-T1 (U-Net)} & 0.878          & 0.845          & 0.835          & 0.814          & 0.785          & 0.725          & 0.604          \\ \hline
\end{tabular}%
}\caption{Average SSIM index on 10 NIND:C500D scenes denoised with models trained on NIND:X-T1 or BM3D}
\vspace{-0.2cm} 
\label{tableXT1onC500D}
\end{table*}

\begin{figure}[!htbp]
\centering
\includegraphics[width=1\linewidth]{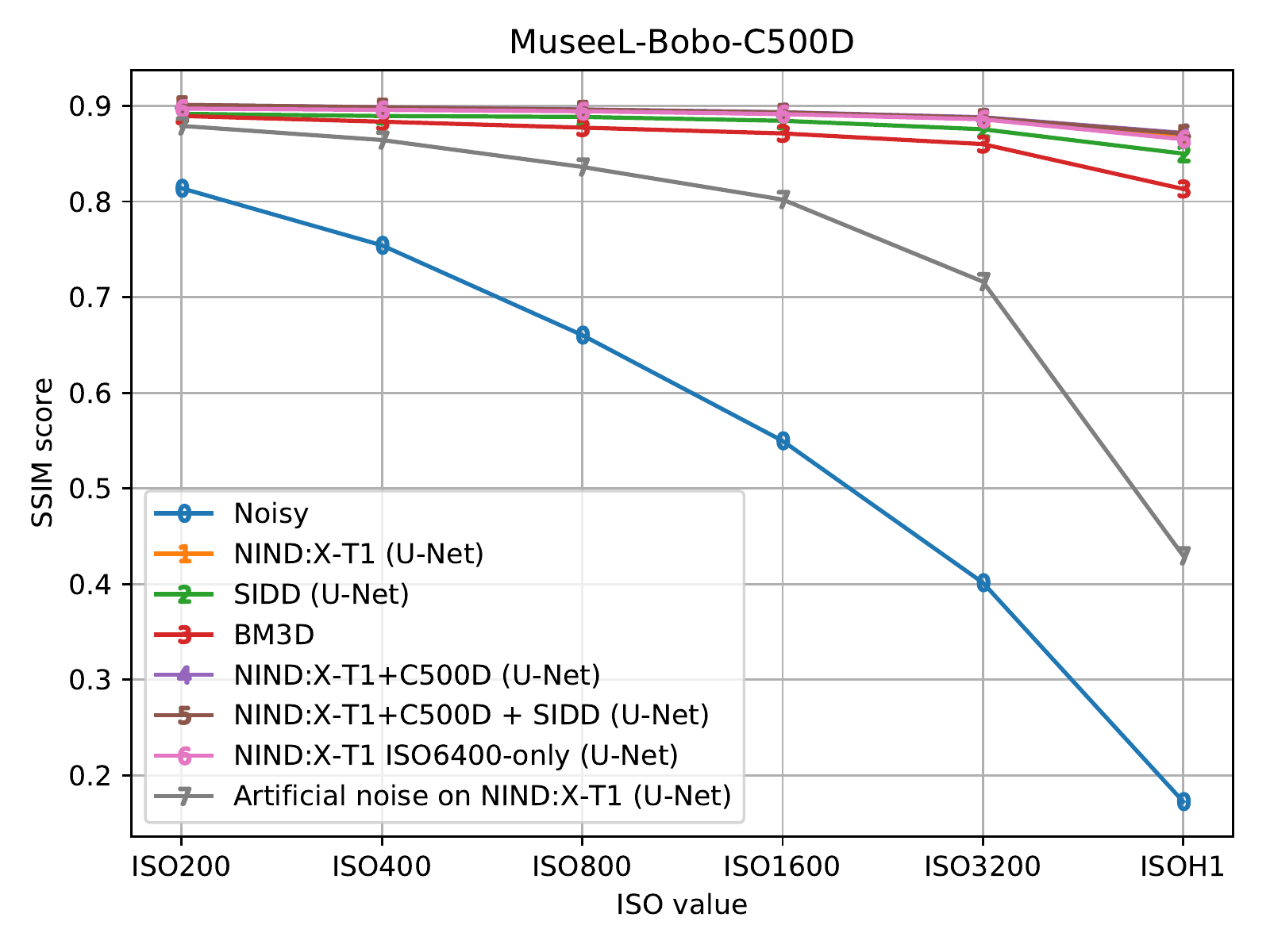}
\caption{Denoising performance of the MuseeL-Bobo-C500D set over increased ISO value (SSIM values shown in Table \ref{tableNINDXT1onBoboC500D})}
\label{fig:boboc500d}
\vspace{-0.2cm} 
\end{figure}

\begin{figure*}[!htbp]
\centering
\includegraphics[width=1\linewidth]{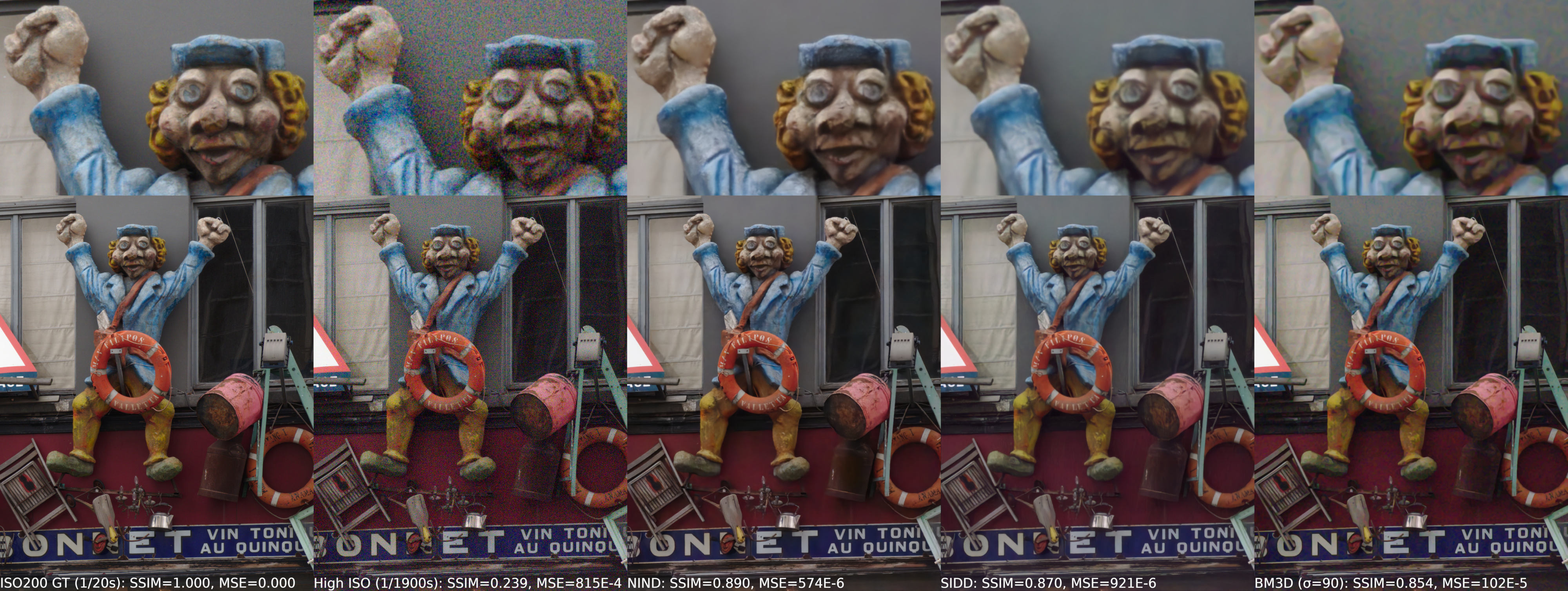}
\caption[caption for stefantiek]{Denoising stefantiek. 1: ISO200 ground-truth (1/20s), 2: high ISO (1/1900s), 3: 2 denoised using U-Net model trained with NIND, 4: 2 denoised using U-Net model trained with SIDD (320-sets), 5: 2 denoised using BM3D ($\sigma=90$\footnoteref{sigmanote})}
\label{fig:stefantiek}
\vspace{-0.2cm} 
\end{figure*}


\begin{figure*}[!htbp]
\centering
\includegraphics[width=1\linewidth]{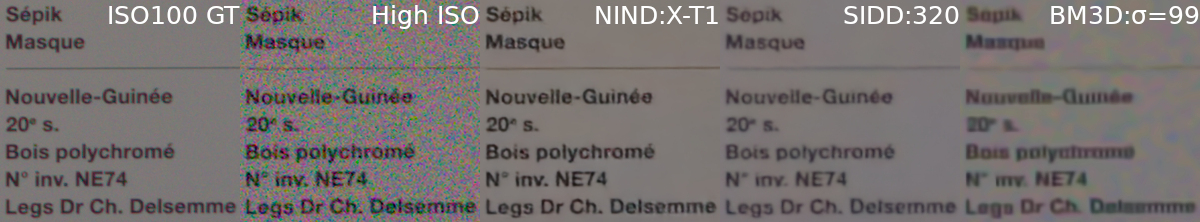}
\caption[caption for sepiktext]{Denoising text present on MuseeL-Sepik-C500D. 1: ISO200 ground-truth (2s), 2: high ISO (1/30s), 3: 2 denoised using U-Net model trained with NIND:X-T1, 4: 2 denoised using U-Net model trained with SIDD (320-sets), 5: 2 denoised using BM3D ($\sigma=99$\footnoteref{sigmanote})}
\label{fig:text}
\vspace{-0.2cm} 
\end{figure*}

\begin{figure}[!htbp]
\centering
\includegraphics[width=1\linewidth]{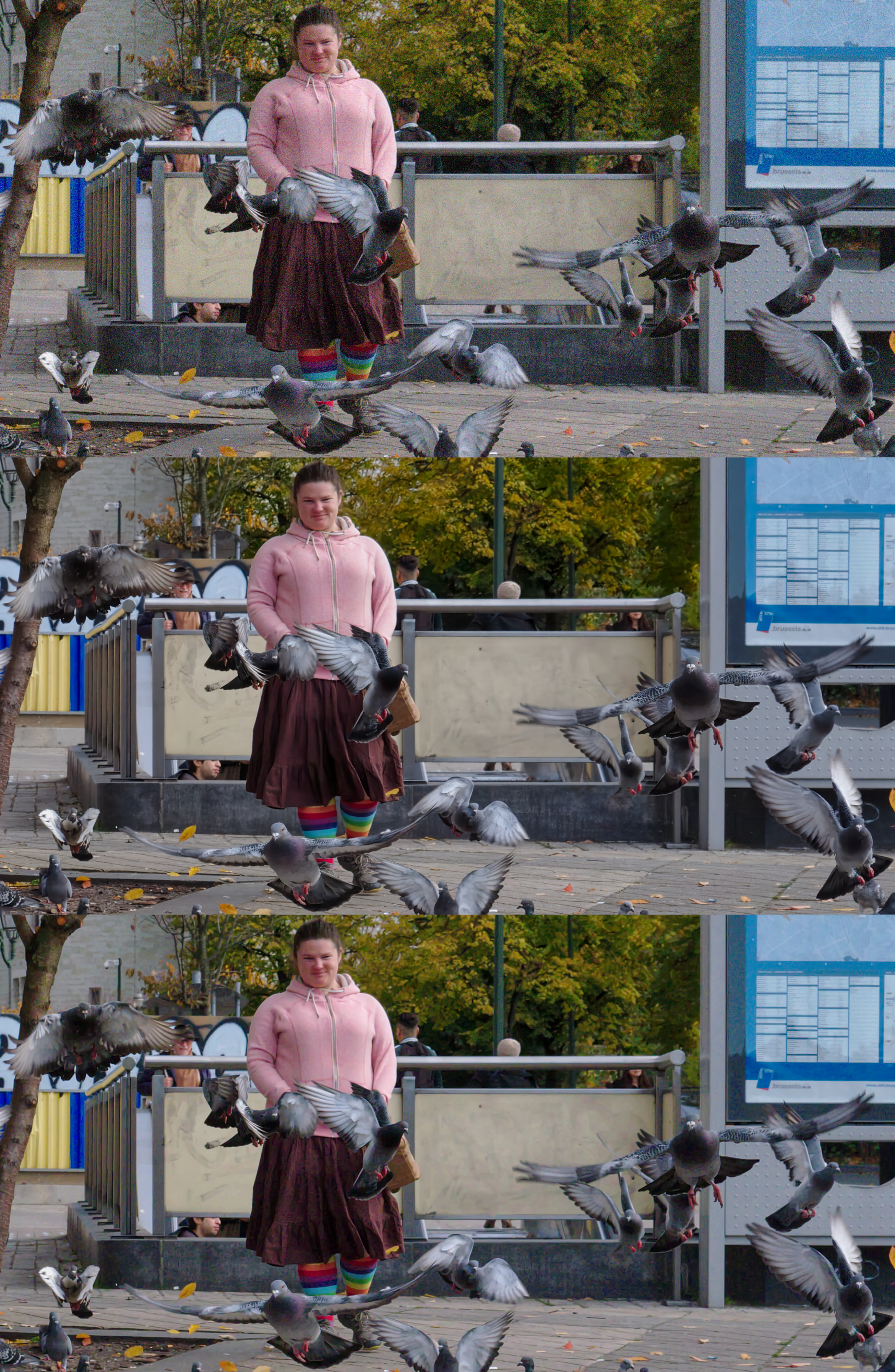}
\caption[pigeoncaption]{Comparison between a noisy ISO6400 crop (top), one denoised with a model trained on NIND (middle), and one on which BM3D ($\sigma=30$) has been applied (bottom). Our model appears to perform well on dynamic scenes despite having been trained on static scenes.}
\label{fig:visualpigeons}
\vspace{-0.5cm} 
\end{figure}

Each network is trained for 48-hours on a GeForce GTX 1080 (11GB). 
Table \ref{tableNINDXT1} shows denoising performance on the Fujifilm X-T1 test pictures. We observe that the model trained on the NIND significantly outperforms BM3D and that adding training data from the Canon EOS 500D sensor, as well as part of the SIDD dataset, does not appear to negatively impact performance. Table \ref{tableNINDXT1onBoboC500D} and Figure \ref{fig:boboc500d} show denoising performance on the scene ``MuseeL-Bobo-C500D", where a model trained only with NIND:X-T1 data performs nearly as well as a model that was also trained with NIND:C500D data (and so does a model trained with both NIND and SIDD). Table \ref{tableXT1onC500D} summarizes the average performance of X-T1-trained models on ten C500D scenes and shows performance which considerably exceeds that of BM3D even though the model was generalizing for a different sensor type.

A model trained with only NIND:X-T1 ISO6400 noisy images yields slightly better performance at and around ISO6400, but this comes with a considerable loss of detail at low ISO and the denoising performance becomes poor as the noise level increases. Moreover the model trained on Fujifilm X-T1 ISO6400 images appears not to generalize as well to different sensors as we found it consistently performs worse on the Canon 500D images. These findings suggest that the cost of generalization is acceptably low and therefore a model mostly benefits from learning with different noise levels and sensors.

Reconstructing the noise (Method \ref{exp:nindmakenoise}) as was suggested in \cite{dncnn} typically yields performance below that of BM3D when applied to ISO noise. The difficulty in reconstructing ISO noise was further noticed in an experiment where we mistakenly fed our learning model noisy images as ground-truth 31\% of the time and it still exceeded BM3D performance. This went as far as inverting the clean and noisy crops and still learning an appreciable level of denoising. These findings suggest that a model can easily tolerate some noise in the ground-truth data. Research into this topic \cite{noise2noise}  has been performed to explicitly train models on noisy data and rely on the zero-mean nature of the noise to effectively remove various artificial noise distributions.

Training a model using artificial noise added to ground-truth images (Method \ref{exp:nindart}), as is commonplace in the literature \cite{rednet}\cite{dncnn}, yields the worst performance in our tests.

In addition to the aforementioned results based on SSIM, we have subjectively tested our NIND-trained model on single images which are not part of the dataset. The first such image is that of a dynamic outdoor scene in which a human walks towards a group of pigeons, causing them to disperse in multiple directions. This type of fast, moving scene cannot be included in the dataset due to its dynamic nature and it must be captured with settings that result in a poor quality image; a small aperture (f/11) to focus everywhere, a fast shutter speed (1/1500s) to capture the flying birds, and a maximum sensor sensitivity (ISO6400) to match the aforementioned settings. Nonetheless, we found the denoised image to be of high quality; we submitted it to the Wikimedia Commons ``Quality Images Candidates" page \cite{qic} and it was subsequently promoted to a ``Quality Image" by Wikimedia Commons reviewers. A crop of the image is provided on Figure \ref{fig:visualpigeons} with a comparison between the noisy version, one denoised with a U-Net model trained on NIND, and a version that has been denoised using BM3D (with $\sigma=30$ which, on average, yields the highest SSIM in our ISO6400 test images). The BM3D version shows significant displeasing artifacts, for example on the skirt and the blue uniform panel on the right, while the model trained on NIND smoothed these regions appropriately while retaining a greater level of useful details such as those present on the pigeons' wings.

\section{Discussion}

We found our model somewhat challenging to use on human subjects. While noise is effectively removed and the level of detail remaining (such as facial hair) is greater than when using conventional denoising methods, human viewers are particularly sensitive to small imperfections (or lack thereof) on human faces and so an overly smooth face may look off-putting. We hope that representative human data would improve the model's performance (much like the purposeful inclusion of text may have helped to reach the domain-specific performance shown in Figure \ref{fig:text}). However, it is difficult to find perfectly still human subjects.

Adding data from different types of sensors should be beneficial as well, as we have seen a slight performance increase when denoising Canon 500D pictures with models trained on both X-T1 and 500D data (rather than 500D-only), yet there was virtually no performance loss on the X-T1 denoised images when we added 500D images to the training data. Likewise, adding the smartphone SIDD dataset \cite{sidd} to a network's training data did not cause any noticeable loss.

Integrating a trained denoiser directly into the pixel-pipeline of image development software (such as darktable \cite{darktable} or GIMP) would be highly beneficial. The noise removal network would be introduced right after the demosaic and exposure steps and thus avoid later noise-amplifying steps.

Most of our experiments are made on single U-Net networks, but modern approaches such as conditional generative adversarial networks (GANs) are likely to yield better performance. pix2pix also uses a U-Net as the generator, but the loss function is replaced with a dedicated ``PatchGAN" discriminator network \cite{pix2pix}. This architecture could be applied to noisy/clean image pairs. BicycleGAN \cite{cyclegan} works in a similar manner but uses a cyclic loss with another generator attempting to generate the original image back, though our results reconstructing the noise (8) seem to indicate that this approach would be less effective. The generative network may as well use a novel architecture such as that proposed in \cite{agan}. GANs benefit from not using a predefined loss function, so they can focus on structured and representative features (such as believable facial features or vital detail in medical imaging \cite{ctgan}) rather than a pixel-loss that is based on a non-existent one-to-one mapping. GANs have also been used to learn and generate noise samples that may be used for training \cite{generatenoiseGAN}; the performance of such an architecture when compared to a model trained on ISO noise remains to be determined. Besides GANs, there are also entirely different types of loss functions which are not pixel-to-pixel based \cite{contextualloss} and may therefore perform better in the denoising domain where cleaning images may introduce blur as there is no one-to-one mapping.

\section{Conclusion}

We released a dataset of photographic ISO noise with scenes captured using multiple ISO values (and matching settings) which can be used to train a blind denoising model. The scenes are static and vary in their content such that the resulting model may adapt to any situation, including dynamic scenes. The dataset currently focuses on large sensors such as those found on DSLR and mirrorless cameras, though we observed no perceptible loss from generalization and therefore expect it may be combined with diverse data such as the SIDD \cite{sidd} that targets smartphones. The Natural Image Noise Dataset is maintained on Wikimedia Commons with the aim of facilitating further contributions and curation.



{\small
\bibliographystyle{ieee}
\bibliography{egpaper_for_review}
}

\end{document}